\documentclass[aps,prl,twocolumn,floatfix,showpacs]{revtex4}

\usepackage{graphicx}
\usepackage{amsmath}
\usepackage{amssymb}
\usepackage{wasysym}

\begin{document}

\title{Correlation between magnetism and spin-dependent transport in CoFeB alloys}

\author{P.~V.~Paluskar}
\email[]{p.v.paluskar@tue.nl}
\author{R.~Lavrijsen}
\author{M.~Sicot}
\author{J.~T.~Kohlhepp}
\author{H.~J.~M.~Swagten}
\author{B.~Koopmans}

\affiliation{Department of Applied Physics, cNM, Eindhoven University of Technology, 5600 MB, The Netherlands}

\date{\today}

\begin{abstract}

We report a correlation between the spin polarization of the tunneling electrons (TSP) and the magnetic moment of amorphous CoFeB alloys. Such a correlation is surprising since the TSP involves \textit{s}-like electrons close to the Fermi level ($E_F$), while the magnetic moment mainly arises due to all \textit{d}-electrons below $E_F$. We show that probing the \textit{s} and \textit{d}-bands individually provides clear and crucial evidence for such a correlation to exist through \textit{s-d} hybridization, and demonstrate the tunability of the electronic and magnetic properties of CoFeB alloys.

\end{abstract}


\maketitle

\begin{figure}[!t]
    \begin{center}
      \includegraphics[width=8.7 cm]{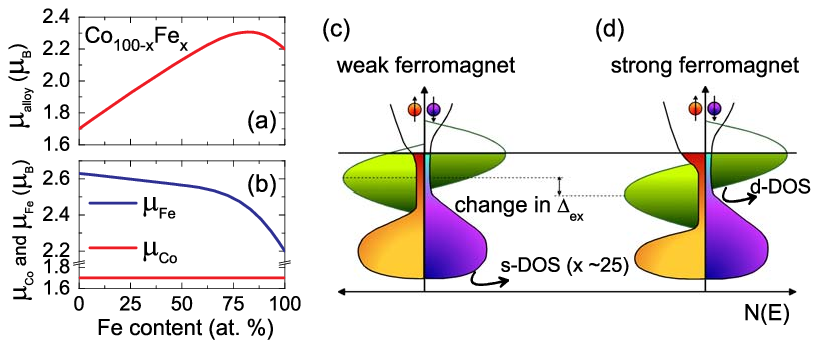}
      \caption{Schematic representation of the Slater-Pauling behavior for Co$_\textrm{{100-x}}$Fe$_\textrm{{x}}$: (a)~S$-$P curve of $\mu_{\textrm{alloy}}$~\cite{RichterPhysicaScripta}. (b)~Known trend of the element-specific Co and Fe magnetic moments~\cite{Collins}. Schematic DOS of (c) weak and (d) strong ferromagnets.}\vspace{-6mm}
      \label{Fig1}
    \end{center}
\end{figure}

At the very foundation of spintronics lie the facts that the conduction electrons in transition metal ferromagnets possess high mobilities and that they get highly spin-polarized as a consequence of their interaction with localized \textit{d}-electrons~\cite{FertReview}. In magnetic tunnel junctions, these \textit{s}-like electrons dominate the tunneling current and are primarily responsible for the tunneling magnetoresistance effect~\cite{YuasaScience,PaluskarPRL}. Early experiments to measure the spin polarization of these tunneling electrons (TSP) in Ni$_{1-\textrm{x}}$Fe$_\textrm{x}$ alloys yielded the unexpected result that the  alloy magnetic moment ($\mu_{\textrm{alloy}}$) \textit{as well as} their TSP displayed the Slater-Pauling (S$-$P) behavior~\cite{MeserveyPRL1976}. The S$-$P behavior of $\mu_{\textrm{alloy}}$ [see Figure~\ref{Fig1}(a)] is the well-known deviation from a linear change resulting in a maximum~\cite{RichterPhysicaScripta,Collins} as the alloy composition changes. While this non-monotonous behavior of $\mu_{\textrm{alloy}}$ is commonly observed in transition metal compounds, their TSP exhibiting a similar curve is very surprising. This surprise stems from the fact that, while $\mu_{\textrm{alloy}}$ is an integral over all states below the Fermi level ($E_F$) and is dominated by \textit{d}-electrons, the TSP originates from transport of \textit{s}-like electrons close to $E_F$. This correlation has been observed only occasionally in experiments~\cite{KaiserPRL94,HindmarchPRB72_100401,KaiserPRL95,MonsmaAPl}. However, the understanding of such a correlation has been neither experimentally nor theoretically addressed, making it a fundamental, long-standing and highly debated issue. Moreover, the existence of such a correlation between $\mu_{\textrm{alloy}}$ and TSP will allow the engineering and tuning of magnetic and electronic properties of ferromagnetic alloys for application in spintronics. We believe that the key to understand this correlation is a combined study of the element-specific \textit{d}-band electronic structure and the \textit{s}-electron dominated TSP in a perceptively chosen material.

In this Letter, we demonstrate the S$-$P behavior of both the TSP and $\mu_{\textrm{alloy}}$ of amorphous Co$_\textrm{{80-x}}$Fe$_\textrm{{x}}$B$_\textrm{{20}}$ alloys. The measured curves of both these properties show distinct similarity in trend and provide an undisputable hint to this correlation. Together with an intuitive understanding of the correlation, we also report a detailed insight in to the various aspects of Co$_\textrm{{80-x}}$Fe$_\textrm{{x}}$B$_\textrm{{20}}$ electronic structure. CoFeB alloys are specifically chosen since: (i) being amorphous, they are highly insensitive to the miscibility of their constituents. (ii) Contrary to most crystalline alloys, their atomic structure does not undergo structural transitions with their composition on the microscopic scale. Both the above distinctions allow easy experimental access to their characteristic properties. (iii) Given their unquestionable importance in spintronics today~\cite{TulapurkarNature,KubotaNatPhys}, and their complex ternary amorphous nature, a comprehensive effort to understand their intrinsic properties remains to be embarked upon.

Since the basic mechanisms for this correlation must involve the electronic structure of the \textit{d}-bands, we use x-ray absorption (XAS) and magnetic circular dichroism (XMCD) to probe their properties. These techniques demonstrate a direct observation of the S$-$P behavior for the orbital ($m_{\textrm{o}}$) and spin ($m_{\textrm{s}}$) moments, as well as the expected changes in the exchange splitting ($\Delta_{\textrm{ex}}$). Together, the observations of the S$-$P behavior of $m_{\textrm{o}}$, and the S$-$P behavior of $m_{\textrm{s}}$ and $\Delta_{\textrm{ex}}$, provide strong evidence to establish that the alteration of the electronic structure with changing alloy composition is, through \textit{s-d} hybridization, primarily responsible for the correlated behavior of $\mu_{\textrm{alloy}}$ and TSP. We would also like to emphasize that such a clear observation of the S$-$P behavior, a characteristic of most transition metal ferromagnetic alloys, has not been established yet using the XMCD technique. Moreover, with this demonstrated tunability and insight into their magnetic, electronic and transport properties, and their low magnetic anisotropy, we believe that CoFeB alloys open several new possibilities to engineer and enhance the performance of spin-torque devices.

We sputter deposited Co$_\textrm{{80-x}}$Fe$_\textrm{{x}}$B$_\textrm{{20}}$ layers from separate targets for each alloy composition. X-ray diffraction (XRD - Cu K$_{\alpha}$) revealed a smooth growth on both SiO$_\textrm{x}$ and AlO$_\textrm{x}$ in an amorphous / nanocrystalline state. $\mu_{\textrm{alloy}}$ was measured using superconducting quantum interference device (SQUID). The TSP data were measured using superconducting tunneling spectroscopy (STS)~\cite{meservey,PaluskarPRL}. Ultraviolet photoemission spectroscopy (UPS) data were measured~\textit{in-situ} at normal emission with a He-I line (21.2~eV). The XAS and XMCD measurements were performed on 120~\AA~CoFeB layers at station 5U.1 of the Daresbury labs by measuring the total electron yield. An external field ($\mu_0H$$\sim$500~mT) was applied at 45$^\circ$ to the photon \textit{k}-vector and the measured spectra were corrected for this angle and photon polarization.

A schematic representation of the S$-$P curve is exemplified for Co$_{100-\textrm{x}}$Fe$_{\textrm{x}}$ alloys in Figure~\ref{Fig1}(a) as a function of the Fe content. Notice that the generic shape for the total magnetic moment is simply a concentration weighted average of element-specific moments of Co and Fe shown in Figure~\ref{Fig1}(b). As sketched in the density of states (DOS) of Figure~\ref{Fig1}(d), Co is a strong ferromagnet with its spin-up \textit{d}-band completely filled. Quite generally, as the alloy composition changes, its electronic structure and its magnetic moment remain unaffected~\cite{RichterPhysicaScripta} [see Figure~\ref{Fig1}(b)]. On the contrary, Fe being weakly ferromagnetic with both spin \textit{d}-bands only partially filled [see Figure~\ref{Fig1}(c)] shows a substantial increase in magnetic moment as the Fe content decreases [see Figure~\ref{Fig1}(b)]. Eventually Fe undergoes a crossover from weak to strong ferromagnetism [see Figure~\ref{Fig1}(c and d)]. Note that this crossover of Fe with the associated increase in the Fe moment essentially causes the S$-$P behavior of $\mu_{\textrm{alloy}}$~\cite{RichterPhysicaScripta,Collins}.

\begin{figure}[!t]
    \begin{center}
      \includegraphics[width=8.7 cm]{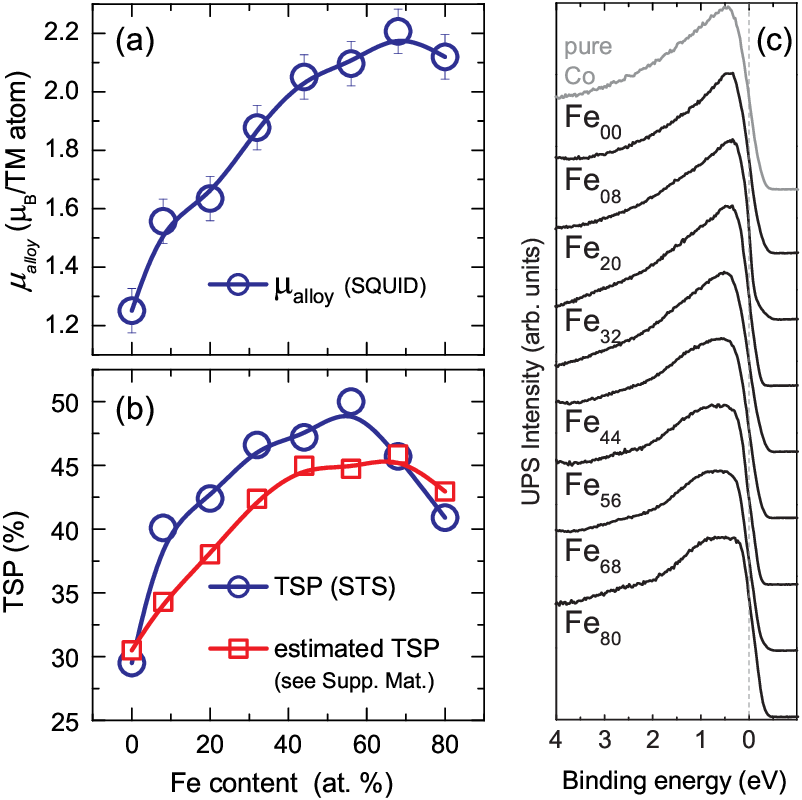}
      \caption{Properties of Co$_\textrm{{80-x}}$Fe$_\textrm{{x}}$B$_\textrm{{20}}$ (a)~$\mu_{\textrm{alloy}}$ measured with SQUID (b)~TSP measured with STS and the estimated TSP (see Supplementary Material and text below). (c)~UPS data.}\vspace{-6mm}
      \label{Fig2}
    \end{center}
\end{figure}

One may ask whether amorphous CoFeB alloys also show the S$-$P behavior. First principles electronic structure calculations predict weak ferromagnetism in amorphous Fe$_\textrm{{80-x}}$B$_\textrm{{x}}$ alloys~\cite{HafnerPRB94} and strong ferromagnetism in amorphous Co$_\textrm{{80-x}}$B$_\textrm{{x}}$ alloys~\cite{TanakaPRB93}. Thus, one may expect that as the Fe content decreases, the Fe DOS undergoes a transition from weak to strong ferromagnetism, which would cause the S$-$P behavior. Just as expected, Figure~\ref{Fig2}(a) shows that $\mu_{\textrm{alloy}}$ of Co$_\textrm{{80-x}}$Fe$_\textrm{{x}}$B$_\textrm{{20}}$ exhibits the S$-$P curve. Such a curve has also been measured for CoFeB before~\cite{O'HandleyAPL}. Next, we focus on their TSP and the changes in their electronic structure which affect it.

The magnitude of the TSP measured as a function of the Fe content is shown as open circles in Figure~\ref{Fig2}(b). Notice that the change in $\mu_{\textrm{alloy}}$ [Figure~\ref{Fig2}(a)] over the whole composition range is around a factor $\sim$$\,$1.7. Remarkably, the TSP too is observed to change by a very similar factor. While the observed correlation in the shape of the two measured curves is not perfect, this similarity between $\mu_{\textrm{alloy}}$ and the TSP is puzzling since, as mentioned earlier, $\mu_{\textrm{alloy}}$ evolves from \textit{d}-electrons while \textit{s}-electrons dominate tunneling through AlO$_\textrm{x}$~\cite{YuasaScience,PaluskarPRL}. Nevertheless, given this apparent correlation, if one naively assumes that the TSP and moment of Co and Fe in the alloy are the same as that in pure Co or Fe films, and that B is unpolarized~\cite{Boron}, then one could estimate the alloy TSP using a simple linear concentration-weighted combination of the known moment and TSP values for pure Co and Fe (see Eqn. S1, Supplementary Material). The TSP values so estimated are shown as open squares ($\Box$) in Figure~\ref{Fig2}(b). One notes a striking similarity of this curve with the measured TSP as well as with $\mu_{\textrm{alloy}}$. In fact, the use of this \textit{crude and admittedly oversimplified approximation} seemingly estimates the alloy TSP within $\sim$5\%~of its measured value. Given this oversimplified approximation, one may wonder whether bulk electronic and magnetic properties may be fit to describe electronic transport at the interface. However, as we have shown in our previous study~\cite{PaluskarPRL}, interface bonding effects at such a complex interface between an amorphous barrier and a chemically and structurally disordered ternary amorphous alloy are an average over the configuration space. In other words, at the interface, (i) the arrangement of each atomic species in the ferromagnet with respect to those of the oxide, and (ii) the variation in the local coordination within the ferromagnetic alloy, are expected to change from site to site. Consequently, though bonding may play a significant role \textit{locally}, the effect of such bonding may average out over a macroscopic junction.

In order to get some insight in the changes of the electronic structure which cause this apparent correlation between TSP and $\mu_{\textrm{alloy}}$, we measured valence band spectra using UPS [see Figure~\ref{Fig2}(c)]. A systematic and pronounced impact of the changing alloy composition on the valence band structure is seen in the spectra. The sharp peak around 0.5~eV for the Co-rich compositions broadens as the Fe content increases up to Fe$_{56}$ and then levels off. Based on the behavior of $\mu_{\textrm{alloy}}$, we tentatively ascribe this pronounced spectral change to the gradual crossover from weak to strong ferromagnetism in the alloys (see Supplementary Material).

The UPS spectra provide a clear and direct evidence on the systematic changes occurring in the electronic structure. However, they are not element-specific. Such an insight would be invaluable considering that the S$-$P behavior essentially derives from the changes in the Fe electronic structure. Therefore, we performed XAS and XMCD at the Fe L$_{2\textrm{,}3}$ edges, probing the Fe \textit{d}-DOS using synchrotron radiation. Next, we will discuss two aspects which can be measured using these techniques: (i) the orbital moment $m_{\textrm{o}}$, and (ii) the spin moment ($m_{\textrm{s}}$) and exchange splitting ($\Delta_{\textrm{ex}}$). The changes in these properties are interrelated. They explicitly demonstrate the transition of Fe from weak to strong ferromagnetism together with the changes occurring in the DOS at $E_F$. Moreover, as we shall see later, this transition also provides a simple picture of a correlation between the \textit{s} and \textit{d}-electrons.

\begin{figure}[!t]
    \begin{center}
      \includegraphics[width=8.7 cm]{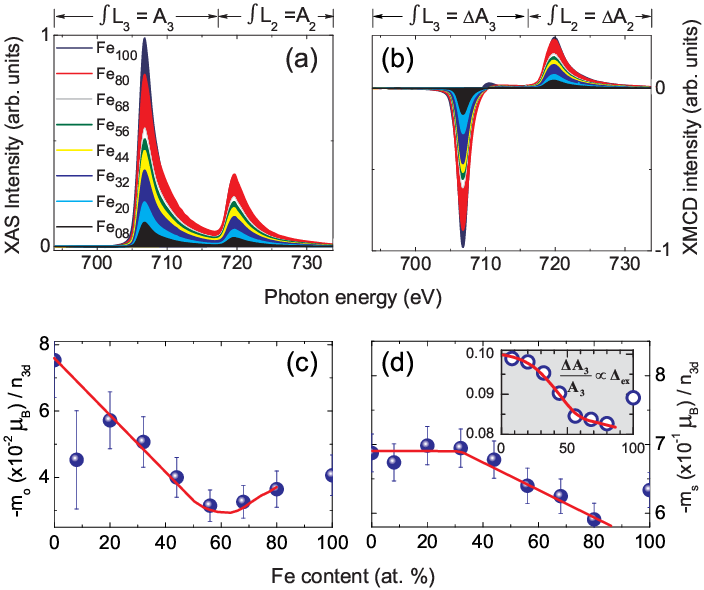}
            \caption{(a)~Background subtracted Fe L$_{2\textrm{,}3}$ edge XAS for Co$_\textrm{{80-x}}$Fe$_\textrm{{x}}$B$_\textrm{{20}}$. (b)~Corresponding XMCD spectra. (c)~Orbital moment per hole, $\frac{m_{\textrm{o}}}{n_{3\textrm{d}}}$. (d)~Spin moment per hole, $\frac{m_{\textrm{s}}}{n_{3\textrm{d}}}$. Inset shows $\frac{\Delta\textrm{A}_3}{\textrm{A}_3}$$\propto$$\Delta_{\textrm{ex}}$~\cite{ChenPRB43_6785}. Fe$_{0}$ represents Co$_{80}$B$_{20}$ measured at the Co L$_{2,3}$ edges. Lines in c and d are guides to the eye.} \vspace{-6mm}
      \label{Fig3}
    \end{center}
\end{figure}

Figure~\ref{Fig3}(a) shows isotropic XAS spectra with standard background subtraction (step function~\cite{ChenPRL}). The difference in the absorption cross-section measured for left / right circularly polarized ($\sim$$\,$66\%) light results in the corresponding XMCD spectra shown in Figure~\ref{Fig3}(b). In Figure~\ref{Fig3}(a-d), note that Fe$_{100}$ represents pure Fe, while Fe$_{0}$ represents Co$_{80}$B$_{20}$ measured at the Co L$_{2,3}$ edges.

\textit{Orbital moment} ($m_{\textrm{o}}$)\textit{:}~According to Thole~\textit{et al.}, $m_{\textrm{o}}$ is given by the orbital sum rule  $\frac{m_\textrm{o}}{n_{3\textrm{d}}}$ = $\frac{4}{3}$$\frac{\Delta\textrm{A}_3+\Delta\textrm{A}_2}{\textrm{A}_3+\textrm{A}_2}$~\cite{TholePRL}. As shown in Figure~\ref{Fig3}(a), the integrated areas under the L$_{2,3}$ edges of isotropic XAS spectra are used to extract A$_{2,3}$, while the corresponding areas under the XMCD spectra are used to extract $\Delta\textrm{A}_{2,3}$ [see Figure~\ref{Fig3}(b)]. $n_{3\textrm{d}}$ denotes the number of \textit{d}-holes, which are unknown in the case of CoFeB. The calculated $\frac{m_\textrm{o}}{n_{3\textrm{d}}}$ is plotted in Figure~\ref{Fig3}(c). Firstly, the absolute value of $m_\textrm{o}$ measured for Fe$_{100}$ ($\sim$$\,$0.13$\,$$\mu_B$ with the known $n_{3\textrm{d}}$$\,$=$\,$3.4) agrees fairly well with the value of $\sim$$\,$0.1$\,$$\mu_B$ calculated including orbital polarization~\cite{Soderlind}. Moreover, the curve in Figure~\ref{Fig3}(c) resembles an inverted S$-$P curve and implies the quenching of $m_\textrm{o}$ with increasing Fe content. We confirmed this quenching of $m_{\textrm{o}}$ by analyzing other ratios known to be sensitive to the spin-orbit interaction (see Supplementary Material). The changes in the Fe electronic structure sketched in Figure~\ref{Fig1}(b-d) may be shown to directly result in the observed quenching of $m_{\textrm{o}}$. It is known that $m_{\textrm{o}}$~$\propto$~[\textit{n}$^\uparrow$($E_F$)~-~\textit{n}$^\downarrow$($E_F$)], where \textit{n}$^{\uparrow\downarrow}$($E_F$) is the spin-resolved total DOS at $E_F$~\cite{EbertJAP67,ErikssonPRB45,Soderlind}. In other words, $m_{\textrm{o}}$ is directly proportional to the ``magnetic'' DOS at $E_F$. A transition from strong to weak ferromagnetism [i.e., from Figure~\ref{Fig1}(d) to~\ref{Fig1}(c)] where the spin-up band moves towards $E_F$ would result in a decrease in [\textit{n}$^\uparrow$($E_F$)~-~\textit{n}$^\downarrow$($E_F$)]. This will consequently result in the quenching of $m_{\textrm{o}}$ we observe. Later we will see that these changes in the ``magnetic'' DOS at $E_F$ may also affect the TSP.

\textit{Spin moment} ($m_{\textrm{s}}$) \textit{and exchange splitting} ($\Delta_{\textrm{ex}}$)\textit{:}~The change in [\textit{n}$^\uparrow$($E_F$)~-~\textit{n}$^\downarrow$($E_F$)] is expected to have a direct effect on $m_{\textrm{s}}$ which constitutes $\gtrsim$$\,$90\% of the total magnetic moment. Figure~\ref{Fig3}(d) shows $\frac{m_{\textrm{s}}}{n_{3\textrm{d}}}$ calculated using the spin sum $\frac{2\Delta\textrm{A}_3-4\Delta\textrm{A}_2}{\textrm{A}_2+\textrm{A}_3}$~-~$\frac{7\langle T_z\rangle}{n_{3\textrm{d}}}$~\cite{CarraPRL}. The magnetic dipole term ($\langle$$\,$$T_z$$\,$$\rangle$) is neglected as its local contributions are expected to cancel out for an amorphous system~\cite{vanderLaanJMMM}. To begin with, the absolute value of $m_{\textrm{s}}$ for Fe$_{100}$ (2.14$\,$$\mu_B$ with $n_{3\textrm{d}}$$\,$=$\,$3.4) is in excellent agreement with the magnetic moment of pure Fe~\cite{ErikssonPRB45,Soderlind}. Most remarkably, the shape of $\frac{m_{\textrm{s}}}{n_{3\textrm{d}}}$ is distinctly similar to that of $\mu_{\textrm{Fe}}$ shown for Co$_\textrm{{100-x}}$Fe$_\textrm{{x}}$ in Figure~\ref{Fig1}(b). Recall that the shape of this curve in CoFe is associated with the transformation of Fe from a weak to a strong ferromagnet. The analogous behavior of $\frac{m_{\textrm{s}}}{n_{3\textrm{d}}}$ in Figure~\ref{Fig3}(d) demonstrates that, as expected, Fe in CoFeB also undergoes a similar transformation. Accompanying this increase in $m_{\textrm{s}}$, another signature of the S$-$P curve would be a similar increase of $\Delta_{\textrm{ex}}$ which has been shown to be directly proportional to $m_{\textrm{s}}$~\cite{HimpselPRL67}. Such an increase in $\Delta_{\textrm{ex}}$ would also endorse our above arguments about the shifting of the \textit{d}-bands [see Figure~\ref{Fig1}(b-d)] which influences the ``magnetic'' DOS at $E_F$ and $m_{\textrm{o}}$. Now, $\Delta_{\textrm{ex}}$ has been shown to be directly proportional to the $\frac{\Delta\textrm{A}_3}{\textrm{A}_3}$ (and $\frac{\Delta\textrm{A}_2}{\textrm{A}_2}$) ratio~\cite{ChenPRB43_6785}. In the inset of Figure~\ref{Fig3}(d), in agreement with the expected increase in $\Delta_{\textrm{ex}}$~$\propto$~$m_{\textrm{s}}$, the $\frac{\Delta\textrm{A}_3}{\textrm{A}_3}$ ratio also increases. Furthermore, quantitatively speaking, in Figure~\ref{Fig1}(b) the Fe moment in Co$_\textrm{{100-x}}$Fe$_\textrm{{x}}$ alloys is seen to increase by $\sim$$\,$23\%, i.e., from the nominal 2.2$\,$$\mu_B$ to $\sim$$\,$2.6$\,$$\mu_B$. Remarkably, in CoFeB, $m_{\textrm{s}}$ and $\Delta_{\textrm{ex}}$~$\propto$~$\frac{\Delta\textrm{A}_3}{\textrm{A}_3}$ also increase by $\sim$$\,$20\% and $\sim$$\,$25\%, respectively [see Figure~\ref{Fig3}(d)]. Similar to the increase in $\frac{\Delta\textrm{A}_3}{\textrm{A}_3}$, we observe an increase in the $\frac{\Delta\textrm{A}_2}{\textrm{A}_2}$ ratio which is also proportional to $\Delta_{\textrm{ex}}$ (not shown). The absolute numbers for these ratios are also in very good agreement with those calculated by Chen~\textit{et al.}~\cite{ChenPRB43_6785}.

Given this crossover of Fe from weak to strong ferromagnetism, we will now address how exactly these changes in the Fe \textit{d}-bands bring about the S$-$P behavior of the \textit{s}-electron dominated TSP. A clear indication comes from two independent arguments:

(i)~Isomer shifts essentially probe the changes in the \textit{s}-electron charge density at the nucleus. In amorphous Co$_\textrm{{80-x}}$Fe$_\textrm{{x}}$B$_\textrm{{20}}$ these isomer shifts also exhibit the S$-$P behavior~\cite{Orue} due to \textit{s-d} hybridization. Although these measured changes in the charge density represent all \textit{s}-electrons below $E_F$ and are not spin-resolved, they directly point to the interplay between \textit{s} and \textit{d}-electrons.

(ii)~The spin-resolved information is observed in our measurements where the S$-$P like changes in $m_{\textrm{o}}$, $m_{\textrm{s}}$ and $\Delta_{\textrm{ex}}$ provide a direct insight in the underlying mechanism which causes a change in the TSP. More specifically, it is well-known that, due to \textit{s-d} hybridization, the \textit{s}-DOS is suppressed in regions of large \textit{d}-DOS~\cite{PaluskarPRL}~[see sketch in Figure~\ref{Fig1}(c-d)]. As the Fe \textit{d}-bands crossover from weak to strong ferromagnetism, the spin-up \textit{d}-band gradually moves below $E_F$. Recall that this shift in the \textit{d}-band also resulted in the quenching of $m_{\textrm{o}}$~$\propto$~[\textit{n}$^\uparrow$($E_F$)~-~\textit{n}$^\downarrow$($E_F$)].  As shown in Figure~\ref{Fig1}(c), due to this shift in the \textit{d}-bands, one may also imagine an associated increase in the spin-up \textit{s}-electron DOS at $E_F$ [$n^\uparrow_s$($E_F$)]. This consequently increases the spin polarization of the Fe \textit{s}-electrons defined as P$^{\textrm{Fe}}_s$~=~$\frac{n^\uparrow_s(\textrm{E}_\textrm{F})-n^\downarrow_s(\textrm{E}_\textrm{F})}{n^\uparrow_s(\textrm{E}_\textrm{F})+n^\downarrow_s(\textrm{E}_\textrm{F})}$. As a result, P$^{\textrm{Fe}}_s$ behaves in a manner similar to the magnetic moment of Fe in Figure~\ref{Fig1}(b). The alloy spin polarization (P$^{\textrm{alloy}}_s$) will consequently show the S$-$P behavior. Note that this increase in P$^{\textrm{alloy}}_s$ will result in a corresponding increase in TSP, since the TSP is a good representative of P$^{\textrm{alloy}}_s$ for these amorphous ferromagnets~\cite{PaluskarPRL}. Here we assume that P$^{\textrm{Co}}_s$  remains unchanged just like the Co moment in Figure~\ref{Fig1}(b). We verified that the Co moment indeed remains unchanged using Co edge XMCD (see Fig S6, Supplementary Material).

Given this information on the various aspects of CoFeB electronic structure and the coherent picture for the existence of a correlation between $\mu_{\textrm{alloy}}$ and TSP, the discrepancy with the TSP measurements on Co$_\textrm{{100-x}}$Fe$_\textrm{{x}}$ alloys complied from various reports in literature, which do not seem to exhibit the S$-$P behavior, may seem particularly puzzling. However, these alloys are crystalline and are known to undergo structural transitions (bcc$\leftrightarrow$fcc) depending on their compositions, which affect their electronic structure and may obscure a clear insight. Moreover, no composition dependent study which directly focuses on the structure, magnetism and TSP of Co$_\textrm{{100-x}}$Fe$_\textrm{{x}}$ alloys has been reported, nor any detailed XMCD measurements, which appear to be indispensable to address this issue, have been performed. On the contrary, the TSP of Co and Fe alloyed with Ru and V~\cite{KaiserPRL94} is known to exhibit a correlation with $\mu_{\textrm{alloy}}$.~XMCD measurements on these alloys could provide more understanding.

In summary, we investigated the magnetism and TSP of amorphous Co$_\textrm{{80-x}}$Fe$_\textrm{{x}}$B$_\textrm{{20}}$ films. We find that the S$-$P behavior of the alloy magnetic moment is also seen in the \textit{s}-electron dominated TSP. XMCD measurements show a crossover from weak to strong ferromagnetism in the Fe-DOS. To the best of our knowledge, this is the first observation of the S$-$P behavior in transition metal alloys using the XMCD technique. We conclude that this crossover in the Fe-DOS, together with \textit{s-d} hybridization, provides an intuitive understanding of the direct correlation between $\mu_{\textrm{alloy}}$ and TSP.

This work is supported by NanoNed, a Dutch nanotechnology program of the Ministry of Economic Affairs, and by STW-VICI grants. We thank the beam line staff of station 5U.1 at Daresbury labs, particularly Dr. T. Johal, for technical support.

\vspace{5 mm}
\textbf{\textit{For references, please see the section after the Supplementary Material given below.}}
\vspace{5 mm}

\clearpage

\begin{center}
\textbf{SUPPLEMENTARY MATERIAL}

\vspace{5 mm}
\textbf{Correlation between magnetism and spin-dependent transport in CoFeB alloys}
\vspace{5 mm}

\textit{P.V. Paluskar, R. Lavrijsen, M. Sicot,}

\textit{J.T. Kohlhepp, H.J.M. Swagten, and B. Koopmans}
\end{center}

\vspace{7 mm}
\noindent\textbf{1.~~~Simple phenomenological estimate of the TSP}
\vspace{5 mm}

As mentioned in the manuscript, the well known literature based magnetic moment and TSP of pure Co and Fe are used to estimate the alloy TSP. More specifically, we have used the following phenomenological equation to make a simple estimate of the TSP:
\vspace{-7mm}
\begin{center}
    \begin{equation}
    \textrm{TSP}=\mu_{\textrm{alloy}} \times \frac{(80-\textrm{x})~.~\textrm{TSP}_{\textrm{Co}}^\textrm{pure}~+~\textrm{x}~.~\textrm{TSP}_{\textrm{Fe}}^\textrm{pure}} {(80-\textrm{x})~.~\mu_{\textrm{Co}}^\textrm{pure}~+~\textrm{x}~.~\mu_{\textrm{Fe}}^\textrm{pure}} \label{SPnTSP2}
    \end{equation}
\end{center}
Here, TSP$_{\textrm{Co}}^\textrm{pure}\,$=$\,$42\%~and TSP$_{\textrm{Fe}}^\textrm{pure}\,$=$\,$45\%~\cite{MonsmaAPl}, while $\mu_{\textrm{Co}}^\textrm{pure}\,$=$\,$1.7$\,$$\mu_B$~and $\mu_{\textrm{Fe}}^\textrm{pure}\,$=$\,$2.2$\,$$\mu_B$~\cite{RichterPhysicaScripta,ErikssonPRB45,Soderlind}, $\mu_{\textrm{alloy}}$ is the measured alloy moment [see Figure 2(a)] and $x$ is the atomic \% of Fe content in CoFeB. We would like to emphasizes that this a crude, and admittedly oversimplified approximation and is included here only to indicate the apparent relation between the alloy magnetic moment and the alloy TSP which seemingly estimates the TSP within 5\% of the actual measured value.

\vspace{5 mm}
\noindent\textbf{2.~~~Tunable coercivity of CoFeB alloys}
\vspace{5 mm}

\begin{figure}[!t]
    \begin{center}
      \includegraphics[width=8.7 cm]{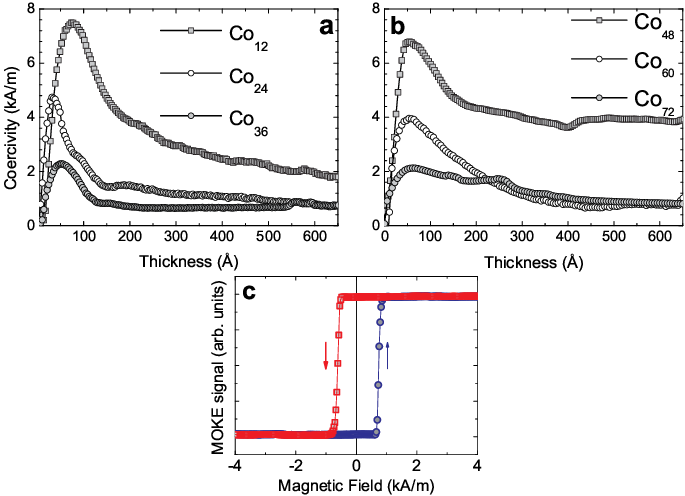} \vspace{-2 mm}
      \caption{(a) and (b) The coercivity of CoFeB layers plotted as a function of film thickness. (c) shows a representative MOKE loop from which the coercivity is extracted.} \vspace{-6 mm}
      \label{Supp6}
    \end{center}
\end{figure}

In Figure~\ref{Supp6}(a \& b) we plot the coercivity of CoFeB layers plotted as a function of film thickness. These measurements were done on wedge shaped samples and probed with magneto-optical Kerr effect (MOKE). In Figure~\ref{Supp6}(c) a representative MOKE loop is shown from which the coercivity is extracted. Although, a complete analysis of the curves in Figure~\ref{Supp6}(a \& b) is beyond the scope of this article, it is clear that as the film composition and thickness are changed, there is a strong variation in the coercivity of these amorphous ferromagnets.

\vspace{5mm}
\noindent\textbf{3.~~~Possible origin of the S-P behavior of $\mu_{\textrm{alloy}}$ in amorphous CoFeB}
\vspace{5mm}
\newline\indent As pointed out in the manuscript, Fe being weakly ferromagnetic with both spin \textit{d}-bands only partially filled shows a substantial increase in its magnetic moment as the Fe content decreases, resulting in the S$-$P curve of $\mu_{\textrm{alloy}}$ in CoFe alloys. According to self-consistent density functional calculations of Schwarz \textit{et al.}, this increase in Fe magnetic moment is due to a rising number of Co nearest neighbors, where Fe atoms having \textit{no} Fe nearest neighbors exhibit the largest magnetic moment~\cite{RichterPhysicaScripta}. In the case of amorphous CoFeB, a clue to the underlying mechanism for this S$-$P behaviour comes from extended x-ray absorption fine structure (EXAFS) measurements~\cite{Orue}. Orue \textit{et al.} observe that as the Co content increases, the short range order around the Fe atoms also increases, predominantly due to the rising number of Co nearest neighbors. Similar to CoFe alloys, and in accordance to the calculations of Schwarz~\textit{et al}~\cite{RichterPhysicaScripta}, one may infer that this rise in the number of Co neighbors around Fe leads to increase in the Fe moment and to the S$-$P behavior of $\mu_{\textrm{alloy}}$ of CoFeB. This argument is substantiated by first-principle calculations on amorphous Co-rich Co$_\textrm{{72}}$Fe$_\textrm{{20}}$B$_\textrm{{8}}$ where Fe is observed to be in a strong ferromagnetic state~\cite{PaluskarPRL}.

\vspace{5 mm}
\noindent\textbf{4.~~~Difference between Fe and Fe$_{80}$B$_{20}$ - XAS}
\vspace{5 mm}

\begin{figure}[!t]
    \begin{center}
      \includegraphics[width=6.0 cm]{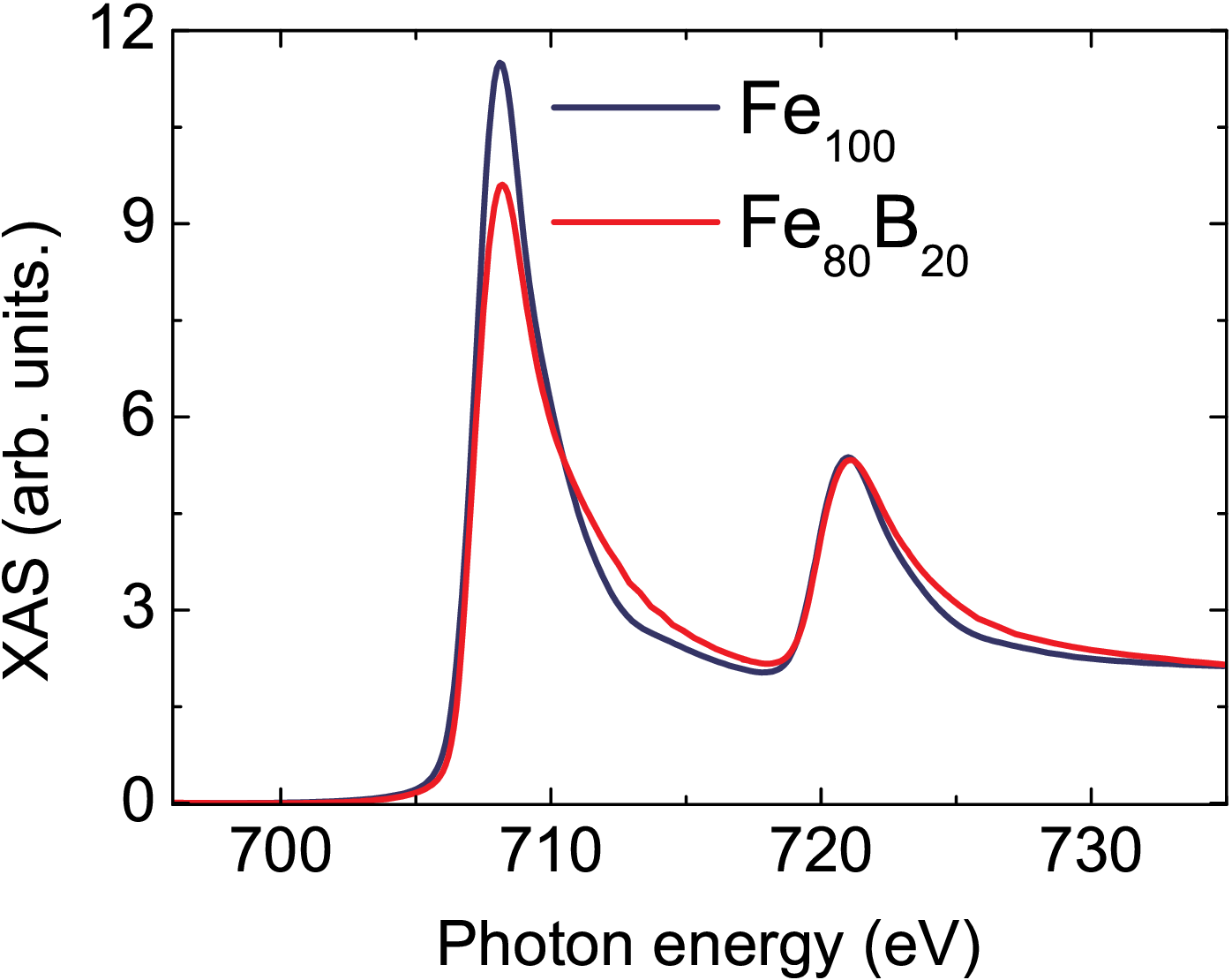} \vspace{-2 mm}
      \caption{Comparison of absorption cross-section ($\Gamma$) data for pure crystalline Fe to that of amorphous Fe$_{80}$B$_{20}$.} \vspace{-6 mm}
      \label{Supp1}
    \end{center}
\end{figure}

Dipole selection rules dictate that an overwhelming majority of transitions are from the L$_2$$\rightarrow$3$d_{3/2}$ final state, and from the L$_3$$\rightarrow$3$d_{5/2}$ final state~\cite{EbertPRBCu}. In other words, the integrals over the L$_{2}$ and L$_{3}$ edges of the isotropic XAS spectra, [A$_2$~and~A$_3$] directly map the unoccupied 3$d_{3/2}$ and 3$d_{5/2}$~DOS, respectively~\cite{EbertPRBCu}. This ability of XAS to probe the nature of the final states is illustrated in Figure~\ref{Supp1} which compares absorption cross-section ($\Gamma$) data for pure crystalline Fe to that of amorphous Fe$_{80}$B$_{20}$. While $\Gamma_{\textrm{L}_2}$ remains unchanged, $\Gamma_{\textrm{L}_3}$ which probes \textit{d}-states higher in the band is seen to decrease for amorphous Fe$_{80}$B$_{20}$. According to electronic structure calculations~\cite{HafnerPRB94}, the exchange splitting in Fe$_{80}$B$_{20}$ is $\sim$0.6~eV smaller in comparison to that of Fe~\cite{HafnerPRB94}. This results in increased occupation of the states higher in the Fe$_{80}$B$_{20}$ \textit{d}-DOS, which may directly lead to a decrease in the absorption ($\Gamma_{\textrm{L}_3}$) to these states. However, the absorption to $\textrm{L}_2$, which probes low-lying states remains largely unchanged~\cite{HafnerPRB94}. The lower value of $\frac{\Delta\textrm{A}_3}{\textrm{A}_3}$~$\propto$~$\Delta_{\textrm{ex}}$ for Fe$_{80}$B$_{20}$ seen in the inset of Figure~3(d) is in good agreement with the lower exchange splitting expected for Fe$_{80}$B$_{20}$ from the above argument. So is the lower value for $\frac{m_{\textrm{s}}}{n_{3\textrm{d}}}$ in Figure~3(d) of the manuscript.

\vspace{5 mm}
\noindent\textbf{5.~~~Band-Filling and orbital moment}
\vspace{5 mm}

\begin{figure}[!t]
    \begin{center}
      \includegraphics[width=6.0 cm]{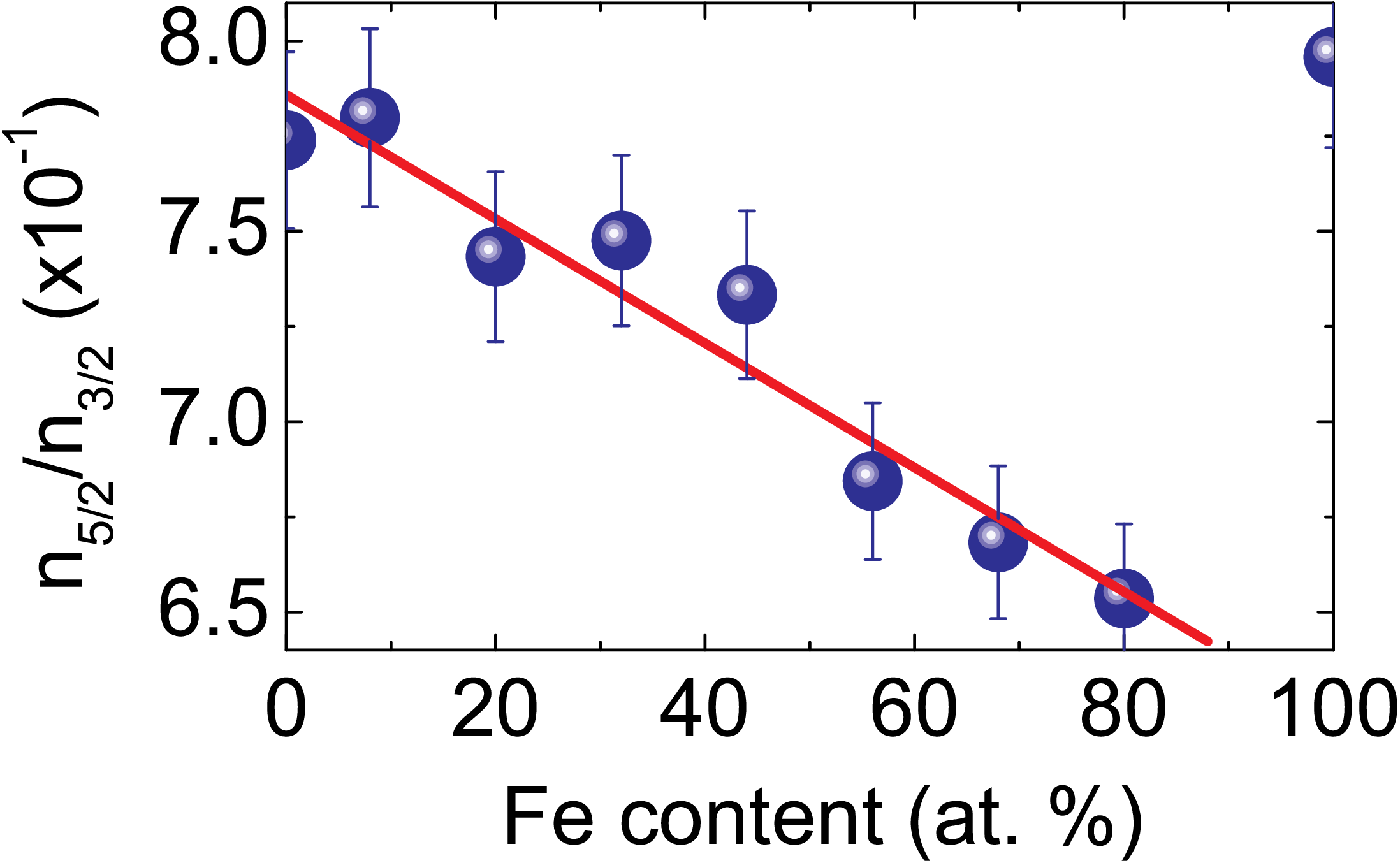} \vspace{-2 mm}
      \caption{The $\frac{n_{5/2}}{n_{3/2}}$ ratio, i.e., the ratio of the occupation of 3d$_{5/2}$ to 3d$_{3/2}$ states.} \vspace{-6 mm}
      \label{Supp2}
    \end{center}
\end{figure}

The orbital moment ($m_{\textrm{o}}$) depends on band-filling effects, the spin moment ($m_{\textrm{s}}$), and short-range order which influences the crystal-field splitting~\cite{Soderlind,ErikssonPRB45}. Band-filling effects can also be studied using XAS. As Fe has one electron less than Co, with increasing Fe content, the gradual removal of one electron can be expected to influence the relative occupancy of the 3$d_{3/2}$ and 3$d_{5/2}$ states. Due to the relatively higher energy of the 3$d_{5/2}$ states, a preferential decrease in their occupancy is expected. This can be analyzed using the $\frac{\textrm{A}_3}{\textrm{A}_2}$ ratio, wherein $\frac{n_{5/2}}{n_{3/2}} = \left(\frac{4.909~\textrm{A}_3}{12~\textrm{A}_2}-\frac{1}{6}\right)$~\cite{MorrisonPRB32_3107}. Here $n_{5/2}$ and $n_{3/2}$ stands for the number of \textit{d}-holes ($n_{3\textrm{d}}$) in the 3$d_{5/2}$ and 3$d_{3/2}$ states. Figure~\ref{Supp2} shows the $\frac{n_{5/2}}{n_{3/2}}$ ratio. Consistent with an intuitive picture, as the Co content increases adding one electron to the system, the plot for $\frac{n_{5/2}}{n_{3/2}}$ suggests that the weight on the 3$d_{5/2}$ states increases. The higher value of $\frac{n_{5/2}}{n_{3/2}}$ for Fe$_{100}$ as compared to Fe$_{80}$B$_{20}$, is in accordance with expected changes in the band-structure and the exchange splitting mentioned above.

\vspace{5 mm}
\noindent\textbf{6.~~~Orbital moment}
\vspace{5 mm}

\begin{figure}[b]
    \begin{center}
      \includegraphics[width=6.0 cm]{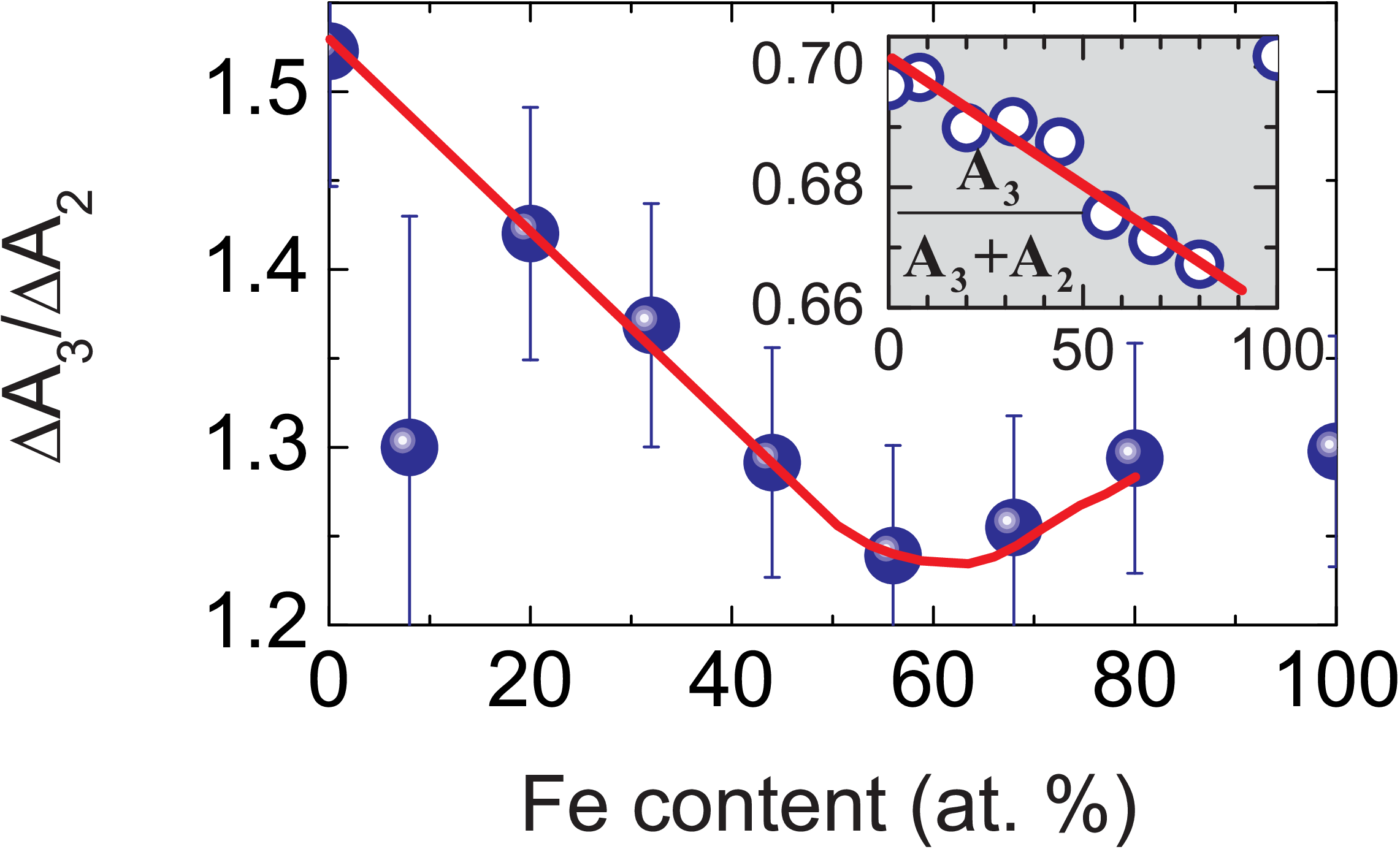} \vspace{-2 mm}
      \caption{Inset shows the $\frac{\textrm{A}_3}{\textrm{A}_3+\textrm{A}_2}$ ratio while the main figure shows the $\frac{\Delta\textrm{A}_3}{\Delta\textrm{A}_2}$~$\propto$~$\xi$~$\propto$~$\frac{m_\textrm{o}}{n_{3\textrm{d}}}$.} \vspace{-6 mm}
      \label{Supp3}
    \end{center}
\end{figure}

The orbital moment can also be probed by using the branching ratio $\frac{\textrm{A}_3}{\textrm{A}_3+\textrm{A}_2}$~\cite{vanderLaanPRL60,TholePRB38}. This calculated ratio is shown in the inset of Figure~\ref{Supp3}. Though the absolute value of the ratio is close to the expected statistical value of 0.66~\cite{vanderLaanPRL60,TholePRB38}, it too shows a decrease with increasing Fe content indicating the quenching of the orbital moment. However, the branching ratio which is derived from XAS is more susceptible to background which arises due to transitions into the continuum. In general, the XMCD spectra are less prone to these issues as they inherently subtract the absorption to the continuum for left and right helicity of the light. Chen \textit{et al.} used relativistic tight-binding calculations to show that the $\frac{\Delta\textrm{A}_3}{\Delta\textrm{A}_2}$ ratio derived from XMCD is very sensitive to the spin-orbit parameter ($\xi$)~\cite{ChenPRB43_6785}. This calculated ratio is shown in Figure~\ref{Supp3}. It too shows the quenching of $\xi$~$\propto$~$m_\textrm{o}$ very similar to the behavior of $\frac{m_\textrm{o}}{n_{3\textrm{d}}}$ in Figure~3(c) of the manuscript. Note that the Fe$_8$ data point is off in Figure~3(c) in the manuscript and in Figure~\ref{Supp3} primarily due to low signal to noise at this low Fe content.

\textbf{Information on the type of short-range order:} S\"{o}derlind~\textit{et al.} calculated that with increasing Fe content, $m_{\textrm{o}}$ decreased if Co$_\textrm{{100-x}}$Fe$_\textrm{{x}}$ was BCC structured~\cite{Soderlind}. This suggests a BCC like short range order for amorphous CoFeB. Interestingly, first-principles atomic structure calculations and EXAFS on amorphous Co$_\textrm{{72}}$Fe$_\textrm{{20}}$B$_\textrm{{8}}$ also showed a BCC-like short range order~\cite{PaluskarPRL,PaluskarUnPub}, contrary to the FCC/densely packed structure expected for such a Co rich alloy.

\vspace{5 mm}
\noindent\textbf{7.~~~Ratio of Orbital to Spin Moment}
\vspace{5 mm}

\begin{figure}[t]
    \begin{center}
      \includegraphics[width=6.0 cm]{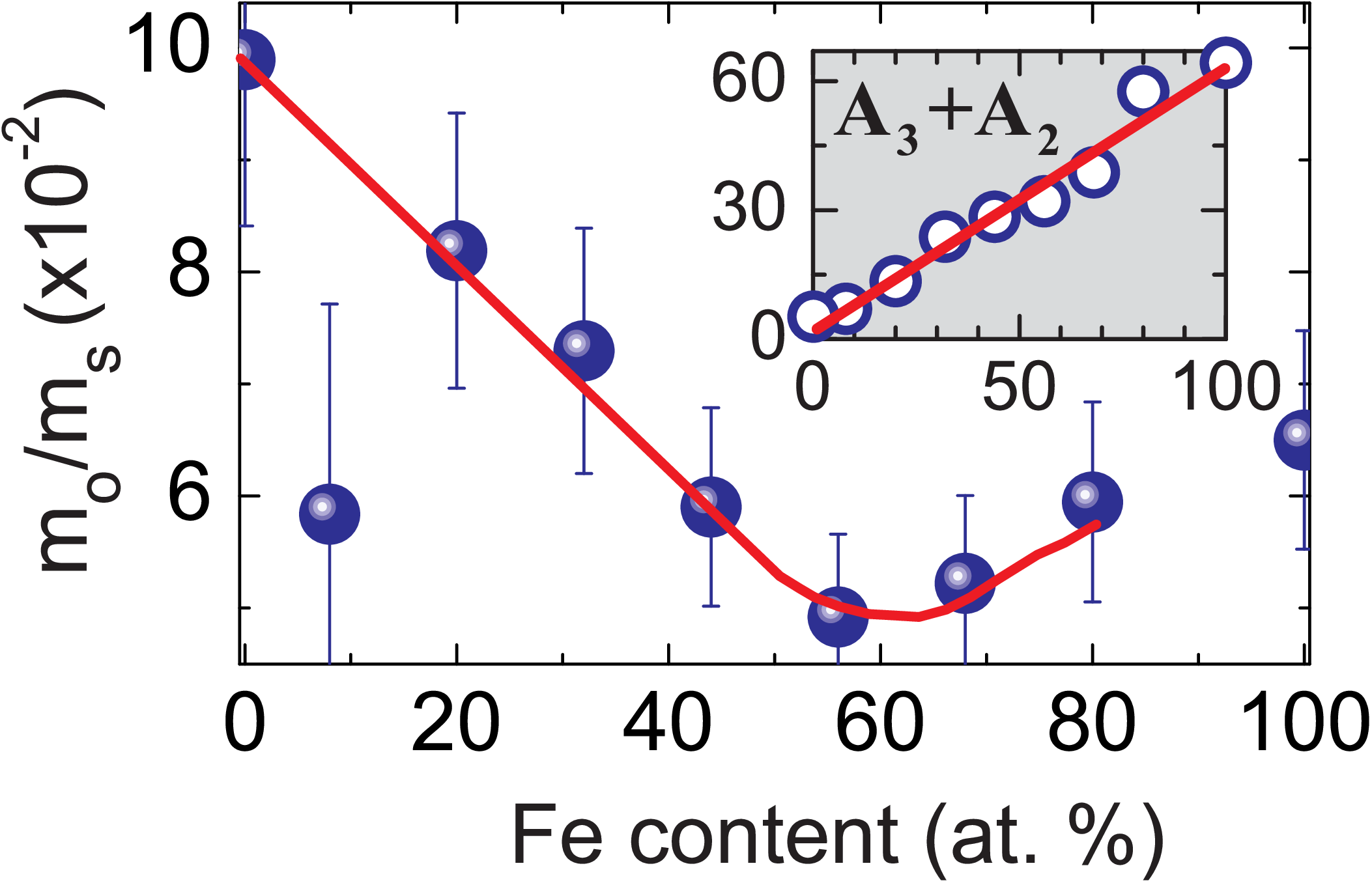} \vspace{-2 mm}
      \caption{Inset shows the A$_3$+A$_2$ also known as the \textit{r} value and associated to the number of holes. The main figure shows the $\frac{m_\textrm{o}}{m_\textrm{s}}$ which is independent of $n_{3\textrm{d}}$~\cite{WuPRL73}. The values for pure Fe and Co films are 4.3 and 9.5 ($\times$10$^-2$) respectively.} \vspace{-6 mm}
      \label{Supp4}
    \end{center}
\end{figure}

The ratio of $\frac{m_{\textrm{o}}}{m_{\textrm{s}}}$~=~$\frac{2}{3}$$\frac{\Delta\textrm{A}_3+\Delta\textrm{A}_2}{\Delta\textrm{A}_3-2\Delta\textrm{A}_2}$ is independent of $n_{3\textrm{d}}$~\cite{WuPRL73}. Note that $n_{3\textrm{d}}$ is unknown for amorphous CoFeB as there are no electronic structure calculations on this alloy series. Chen~\textit{et al.}~\cite{ChenPRL} found $\frac{m_{\textrm{o}}}{m_{\textrm{s}}}$ to be 0.043 for pure BCC Fe and 0.095 for pure FCC Co. Our measurements [see Figure~\ref{Supp4}] on amorphous Co$_\textrm{{80-x}}$Fe$_\textrm{{x}}$B$_\textrm{{20}}$ are in excellent agreement with the work of Chen~\textit{et al.} on crystalline Co and Fe films. The inset in Figure~\ref{Supp4} shows the sum of the areas under the L$_{2,3}$ edges, generally also known as the \textit{r} value and associated to the number of holes. The linear increase with Fe content indicates that the number of holes per Fe atom does not vary with composition.

\vspace{5 mm}
\noindent\textbf{8.~~~Co edge XAS and XMCD}
\vspace{5 mm}

\begin{figure}[!b]
    \begin{center}
      \includegraphics[width=8.7 cm]{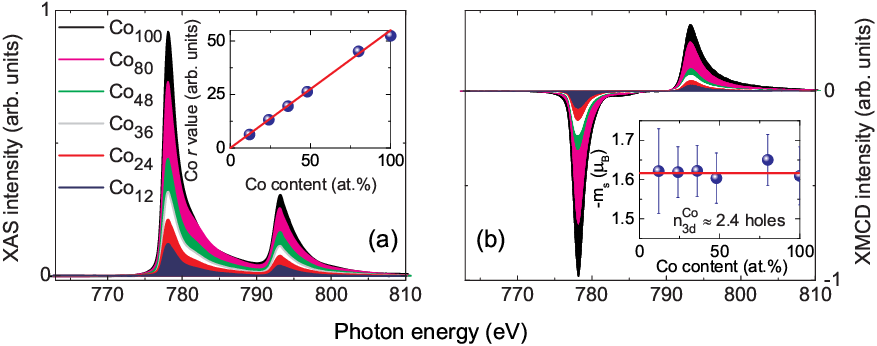} \vspace{-2 mm}
      \caption{\textbf{Co edge XAS and XMCD.} (a) XAS on Co edge. Inset shows A$_3$+A$_2$ also known as the \textit{r} value. (b) XMCD on Co edge. Inset shows the extracted $m_{\textrm{s}}$ where $n_{3\textrm{d}}$ for Co is taken to be the well known value of 2.4 holes~\cite{ErikssonPRB45,Soderlind}.}
      \vspace{-6 mm}
      \label{Supp44}
    \end{center}
\end{figure}

Although limited by the available beam time, we performed XAS and XMCD measurements on the Co edge for most of these alloys. These data are shown for the sake of completeness in Figure~\ref{Supp44}. Similar to the Fe edge, the \textit{r} value (A$_3$+A$_2$) on the Co edge in the inset of Figure~\ref{Supp44}(a) is seen to vary linearly with composition. Regarding the XMCD data shown in in Figure~\ref{Supp44}(b), we observe no change in the spin moment on Co atoms as the composition changes. Here, after evaluation of $\frac{m_\textrm{o}}{n_{3\textrm{d}}}$, the number of holes for Co is taken to be the well known value of 2.4 holes~\cite{ErikssonPRB45,Soderlind}. Recall, that since Co is a strong ferromagnet, one does not expect any changes in its spin magnetic moment, as confirmed by the XMCD data of Figure~\ref{Supp44}(b). Moreover, the obtained value of 1.6$\,$$\mu_B$ for the spin moment of Co is in good agreement with calculations~\cite{ErikssonPRB45,Soderlind}.

\vspace{5 mm}
\noindent\textbf{9.~~~Valance band photoemission (UPS)}
\vspace{5 mm}

As mentioned in the manuscript, in order to get some insight in the changes of the electronic structure, we measured valence band spectra using ultraviolet photoemission spectroscopy (UPS). This technique probes a specific region of the Brillouin zone depending on the energy of the photons and the growth direction of the sample.
\begin{figure}[b]
    \begin{center}
      \includegraphics[height=10.0 cm]{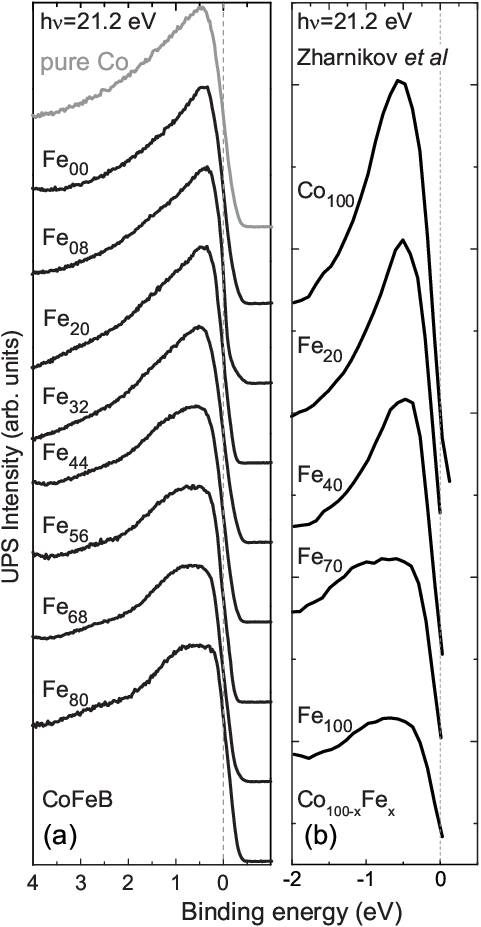} \vspace{-2 mm}
      \caption{(a)~UPS data on Co$_\textrm{{80-x}}$Fe$_\textrm{{x}}$B$_\textrm{{20}}$. (b) UPS data on single crystalline fcc (100) Co$_\textrm{{100-x}}$Fe$_\textrm{{x}}$ alloys. \textit{Data courtesy of Prof.Dr. Wolfgang Kuch~\cite{Zharnikov}}.} \vspace{-6 mm}
      \label{Supp7}
    \end{center}
\end{figure}
Now, it is well known that the UPS spectra of amorphous and single crystalline alloys are very similar to each other~\cite{HasegawaReview,Matsuura,Guntherodt,Amamou}. Based on these previous findings, one may compare our UPS CoFeB data to that on single crystalline (100) Co$_{100-x}$Fe$_{x}$ alloys from Zharnikov \textit{et al.}~\cite{Zharnikov}. In Figure~\ref{Supp7}(a \& b), we compare our data to that of Zharnikov \textit{et al.}~\cite{Zharnikov}. Indeed, one notes that the sharp peak for Co-rich single crystalline Co$_{100-x}$Fe$_{x}$ alloys is similar to amorphous Co$_{80}$B$_{20}$. Moreover, this similarity extends throughout the composition dependent study. By comparing their measurements to semi-relativistic band structure calculations, Zharnikov \textit{et al.} argue that this change of the UPS spectra basically arises from the change in exchange splitting and band filling as the alloy composition is varied~\cite{Zharnikov}. In other words, from the intrinsic difference between the exchange splitting and band filling of Fe and Co electronic structures. Note that this difference in exchange splitting and band filling is the fundamental reason why pure Co is a strong ferromagnet and pure Fe is a weak ferromagnet~\cite{RichterPhysicaScripta}. Therefore, based on the behavior of $\mu_{\textrm{alloy}}$ of our CoFeB alloys and previous measurements of Zharnikov \textit{et al.} on single crystalline CoFe samples, we tentatively ascribe this pronounced valance band spectral change to the gradual crossover from weak to strong ferromagnetism in amorphous CoFeB alloys. As we have seen earlier, XMCD provides clear evidence of the increase in exchange splitting and band filling of these alloys as the composition is varied, endorsing our above arguments.

\vspace{5 mm}
\noindent\textbf{10.~~~TSP measurements}
\vspace{5 mm}

\begin{figure}[!h]
    \begin{center}
      \includegraphics[width=5.5 cm]{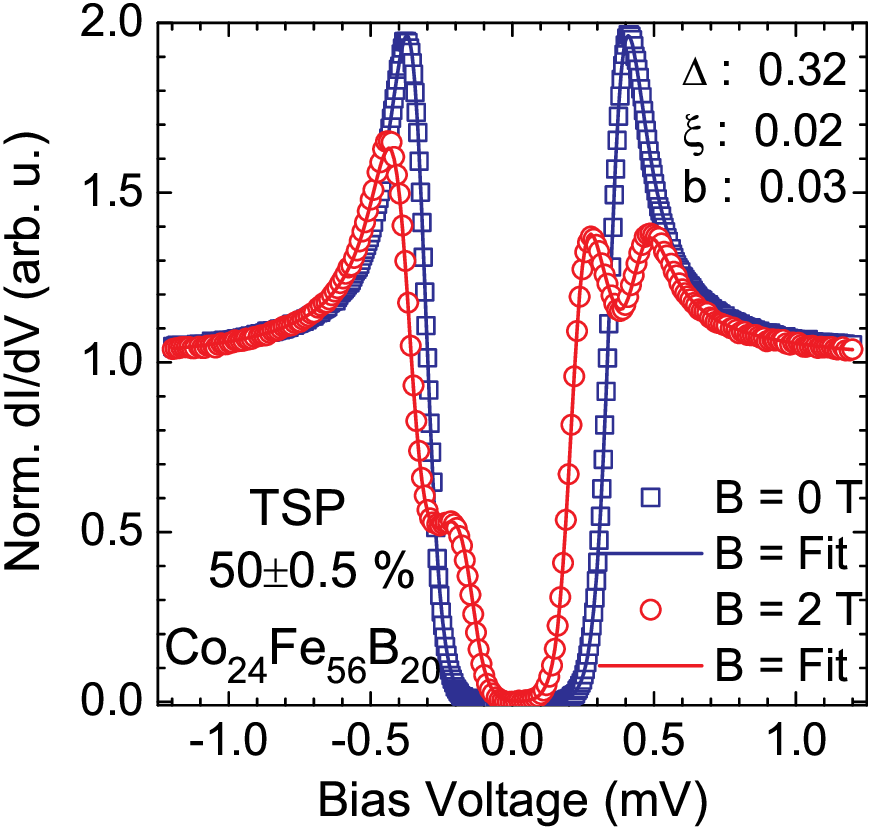} \vspace{-2 mm}
      \caption{Representative superconducting tunneling spectroscopy measurement on Co$_{24}$Fe$_{56}$B$_{20}$. The superconducting gap ($\triangle$), orbital depairing ($\xi$), spin-orbit scattering (\textit{b}) and temperature (T) are fit parameters.} \vspace{-6 mm}
      \label{Supp5}
    \end{center}
\end{figure}

Figure~\ref{Supp5} shows representative TSP data measured at 0.25~K using superconducting tunneling spectroscopy on Al/AlO$_\textrm{x}$/CoFeB/Al junctions. Our junctions show high-quality superconducting gaps with sharp peaks. The zero field curve ($\square$) shows the Al superconducting gap while the application of a magnetic field ($\mu_0$H$\,$$>$$\,$2.0$\,$T) results in the Zeeman-splitting of the Al superconducting DOS which acts as a spin analyzer for the tunneling electrons. The observed asymmetry in the intensity of the measured peaks ($\Circle$) when fit (solid lines) with Maki theory~\cite{maki64} reveals the TSP of Co$_{24}$Fe$_{56}$B$_{20}$. The superconducting gap ($\triangle$), orbital depairing ($\xi$), spin-orbit scattering (\textit{b}) and temperature (T) are fit parameters.

\end{document}